# Memresistors and non-memristive zero-crossing hysteresis curves

Blaise Mouttet

*Abstract*— **It has been erroneously asserted by the circuit theorist Leon Chua that all zero-crossing pinched hysteresis curves define memristors. This claim has been used by Stan Williams of HPLabs to assert that all forms of RRAM and phase change memory are memristors. This paper demonstrates several examples of dynamic systems which fall outside of the constraints of memristive systems and yet also produce the same type of zero-crossing hysteresis curves claimed as a fingerprint for a memristor. This establishes that zero-crossing hysteresis serves as insufficient evidence for a memristor.**

I. INTRODUCTION

Below is a summarized timeline of events so far pertaining to the Chua-HP "memristor":

1) In 1971 Leon Chua publishes a paper proclaiming the *memristor* as the missing fourth fundamental circuit element linking electric charge and magnetic flux linkage [1]. This paper acknowledges the earlier existence of ovonic switches (a precursor to modern phase change memory) but argues that no physical memristor has yet been discovered.

2) In 1976 Leon Chua and Sung-Mo Kang publish a paper introducing *memristive systems* as a special case of state-space dynamic systems [2]. This paper distinguishes memristive systems from the memristor by noting that:

"..*there remains an even broader class of physical devices and systems whose characteristics resemble those of a memristor and yet* **cannot be realistically modeled by this element***,..*"

3) Beginning in the late 1990's and continuing through the 2000's several companies including Micron Technology, Sharp, Samsung, and Unity Semiconductor began experimenting with new forms of 2-terminal resistance memories based on chalcogenide and metal oxide materials. During this period HPLabs was attempting to make molecular switching devices work properly [3]. Also during this period researchers at Samsung invented a new type of resistive memory based on the motion of oxygen ions in $TiO_2$ and other metal oxides [4].

4) In 2008 researchers for HPLabs including Stan Williams claim to have found Chua's missing memristor and simultaneously switch their research efforts away from molecular memory toward $TiO_2$ and other metal oxide resistance switches [5]. The physical structure of $TiO_2$ discussed by the HPLabs researchers is identical to that discussed in the patent application of Samsung [4].

5) In 2010 I gave a presentation at the IEEE International Symposium on Circuits and Systems (ISCAS) noting several problems with the narrative regarding Chua and HP's memristor [6]. During the presentation I noted that most real memory resistor devices could not be accurately modeled using Chua's original 1971 definition of a memristor. I also noted other companies had been working on other forms of RRAM based on other device materials and physical operating mechanisms.

6) In January of 2011 Leon Chua published a paper [7] arguing that:

"*All 2-terminal non-volatile memory devices based on resistance switching are memristors, regardless of the device material and physical operating mechanisms.*"

In order to justify this statement Chua pointed to the 1976 definition of memristive systems rather than the original memristor definition therefore directly contradicting the statement that the broader class of memristive systems cannot be realistically modeled by a memristor.

7) In October of 2011 Stan Williams of HPLabs publically made a claim that phase change memory, MRAM, and RRAM are all memristor technologies [8]. Thus ovonic switches which were originally denied as a form of physical memristor by Chua in 1971 are now suddenly a type of memristor.

There is already evidence in the literature [9] suggesting that Chua's analysis is flawed for failing to consider the distinction between type I (self-crossing pinched) hysteresis curves (usually associated with bipolar memory) and type II (non-self-crossing pinched) hysteresis curves (usually associated with unipolar memory).

The goal of this paper is to further illustrate that it is possible to construct dynamic systems which fall outside of the definition of either memristor or memristive systems and yet still produce zero-crossing pinched hysteresis curves.

## II. EXAMPLES OF NON-MEMRISTIVE DYNAMIC SYSTEMS EXHIBITING ZERO-CROSSING HYSTERESIS

It is first noted that the definition of a memristive systems is broadly given [2] as

$$y(t) = g(x, u, t)u(t)$$
$$\frac{dx(t)}{dt} = f(x, u, t) \quad (1)$$

wherein $u(t)$ = the input signal, $y(t)$ = the output signal, $x(t)$ is the system state (which generally may be a vector function), $g(x,u,t)$ is a **continuous** function most generally dependent on the system state, input signal, and time, and $f(x,u,t)$ is a **continuous** function defining the rate of change of the state dependent on the system state, input signal, and time.

The following examples illustrate that zero-crossing pinched hysteresis curves may be generated by dynamic systems falling outside of the canonical memristive systems framework.

Example 1: 1$^{st}$ order non-linear dynamic system with initial condition x(0)=x$_0$

$$y = u + 2(x - x_0)$$
$$\frac{d(x-x_0)}{dt} = u^2 - \frac{1}{2}A^2 \quad (2)$$

In this case $u$ represents the input signal, $x$ is the state variable, and $y$ is the output signal. *A* may be interpreted as a particular voltage or current level and $A^2/2$ may be interpreted as a particular power level (see example 4 for a related physical system).

For u=Asin(ωt), d(x-x$_0$)/dt = A$^2$(sin$^2$(ωt)-1/2) = -A$^2$cos(2ωt)/2, and x = x$_0$-A$^2$sin(2ωt)/4ω) (given the initial condition x(0)=x$_0$). The output equation can then be expressed as y = Asin(ωt) – A$^2$sin(2ωt)/2ω. The y vs. u hysteresis curve for this case is illustrated in the top figure on page 7 based on a Mathcad plot for ω=1 and A=1.

Example 2: 2$^{nd}$ order non-linear dynamic system with initial conditions x(0)=x$_0$ and dx(0)/dt = -ωA

$$y = u + (x - x_0)\frac{du}{dt}$$
$$\frac{d^2(x-x_0)}{dt^2} = \omega^2 u \quad (3)$$

As in the 1$^{st}$ example, $u$ represents the input signal, $x$ is the state variable, and $y$ is the output signal. For u = Asin(ωt), x-x$_0$ = -Asin(ωt) (given the initial conditions). The output equation can then be expressed as y = Asin(ωt)-ωA$^2$sin(ωt)cos(ωt). The y vs. u hysteresis curve for this case is illustrated in the middle figure on page 7 based on a Mathcad plot for ω=1 and A=1. It is notable that this example may be physically comparable to a memadmittance system in which ionic or oxygen vacancies demonstrate large inertia relative to the damping and electrostatic forces [10,11].

Example 3: Frequency independent pinched hysteresis with x(0)= -A

$$y = u + (A^2 - x^2)x$$
$$\frac{dx}{dt} = \omega u \quad (4)$$

As in the 1$^{st}$ example, $u$ represents the input signal, $x$ is the state variable, and $y$ is the output signal. For u = Asin(ωt), x = -Acos(ωt) (given the initial condition). The output equation can then be expressed as y = Asin(ωt)-A$^3$sin$^2$(ωt)cos(ωt). The y vs. u hysteresis curve for this case is illustrated in the bottom figure of page 7 based on a Mathcad plot for ω=1 and A=1.

In Example 1 it is evident that an increase in the signal frequency produces decay in the hysteresis effect (as is currently assumed to be correct from memristive systems). In Example 2 an increase in the signal frequency increases the hysteresis effect in contrast to memristive systems. Example 3 shows an example of frequency independent pinched hysteresis.

The first three examples prove that it is possible to generate pinched hysteresis using dynamic systems outside of the realm of memristive systems. The next three examples provide some cases with more physical meaning.

Example 4: Thermal memresistor

$$v = Ri + S(T - T_0)$$
$$\frac{d(T-T_0)}{dt} = ki^2 - \alpha(T - T_0) \quad (5)$$

In this case $i$ (electron current) represents the input variable, $T$ (temperature) is the state variable, and $v$ (voltage) is the output variable.

This example is based on an idealized physical system in which a voltage $v$ is derived based on the sum of an ohmic voltage expressed as the product of resistance $R$ and current $i$ and an additional voltage based on the product of the Seebeck coefficient $S$ and a temperature gradient $T-T_0$. In this idealized system even in cases wherein there is no applied current a temperature gradient can induce a voltage. The dynamic equation expresses the rate of variation of the temperature gradient $T-T_0$ based on the difference between the Joule heating ($ki^2$), which raises the temperature of the system, and the dissipated energy $\alpha(T-T_0)$, determining how fast the thermal energy is lost to the environment.

It is evident that this example has a form very similar to that of Example 1 with the exception that a transient component exists in the solution to the dynamic equation. However, it may be feasible to design a system such that, given a current signal i=I$_0$sinωt, the Joule heating in the first term exactly balances the energy loss from the second term. In this case it would be possible to reproduce the same type of hysteresis curve as in Example 1.

It is also evident that this example represents a volatile memory systems but it may be converted to a non-volatile memory system such as phase change memory by including a crystallization rate equation and noting the dependence of the resistance on the crystallization state.

Example 5: Ionic memresistor

$$i = i_d(\Delta x_d, v_a) - i_d(\Delta x_d, 0) \quad (6a)$$

$$\frac{d^2\Delta x_d(t)}{dt^2} + \frac{1}{\tau_c}\frac{d\Delta x_d(t)}{dt} + \frac{(ze)^2 N_d}{m_{ion}\epsilon_r\epsilon_0}\Delta x_d(t) = \left(\frac{a^2 vze}{\tau_c kT}\right) exp\left(\frac{-W_a}{kT}\right)\frac{v_a(t)}{x_{d0}} \quad (6b)$$

A full discussion of the development of this memresistive system is given in [11]. In this case $v_a$ (voltage) is the input variable, $\Delta x_d$ (variation in the depletion width) is the state variable, $i$ (current) is the output variable, and $i_d(\Delta x_d, v_a)$ is a function defining the dependence of the current on the depletion width and the voltage. For the dynamic equation (6b) the ionic doping level is denoted by $N_d$ and the remainder of the terms may be considered constants.

Different versions of the function $i_d(\Delta x_d, v_a)$ may be formulated depending on a particular thin film insulator or semiconductor. In general, however, for a sinusoidal voltage input $v_a$=V$_0$sin(ωt) the output current takes the form

$$i = i_d(\Delta X_d \sin(\omega t + \varphi_0), V_0 \sin \omega t) - i_d(\Delta X_d \sin(\omega t + \varphi_0), 0) \quad (7)$$

where $\Delta X_d$ and $\varphi_0$ are constants depending on the ion doping level, effective ion mass, relative permittivity, temperature, and signal frequency (see equations (39)-(42) of [11]). It is clear from inspection that the i vs. v curve for this system will produce a zero-crossing hysteresis (with the exception of the case $\varphi_0 = 0$ which degenerates the hysteresis).

Example 6: Filamentary memresistor

$$J_T(t) = \begin{array}{c} J_0 exp\left(\frac{-\sqrt{8m_e}}{h/2\pi}\sqrt{\frac{zen_f}{A_f\varepsilon_r\varepsilon_0}}x_f^3(t) - v_a(t)x_f^2(t)\right) \\ -J_0 exp\left(\frac{-\sqrt{8m_e}}{h/2\pi}\sqrt{\frac{zen_f}{A_f\varepsilon_r\varepsilon_0}}x_f^3(t)\right) \end{array} \quad (8a)$$

$$\frac{d^2\Delta x_f(t)}{dt^2} + \frac{1}{\tau_f}\frac{d\Delta x_f(t)}{dt} + \frac{(ze)^2 n_f}{m_f x_{f0} A_f \varepsilon_r \varepsilon_0}\Delta x_f(t) \approx \frac{KzeV_a(t)}{m_f x_{f0}} \quad (8b)$$

$$\Delta x_f(t) = x_f(t) - x_{f0} \quad (8c)$$

A full discussion of the development of this memresistive system is given in [12]. In this case $v_a$ (voltage) is the input variable, $\Delta x_f$ (variation in the filament tunneling gap $x_f$) is the state variable, and $i$ (current) is the output variable. For the dynamic equation (8b) the number of ions associated with the filament is denoted by $n_f$ and the remainder of the terms may be considered constants.

The i vs. v hysteresis curve for some special cases of this type of memresistive system is illustrated on page 6 based on a Mathcad plot for different ion levels (cases 1-3) and based on a phase shift such as resulting from a large effective mass of the filament (case 4).

The next eight examples are representative of generative families of non-memristive dynamic systems which demonstrate zero-crossing hysteresis.

Example 7: 1st order memresistor.

$$y = g(x)\frac{d^2u}{dt^2}$$
$$\frac{dx}{dt} = u \quad (9)$$

As in the 1st example, $u$ represents the input signal, $x$ is the state variable, and $y$ is the output variable with $g(x)$ representing a continuous function of the system state $x$. For $u = A\sin(\omega t)$, $x = x_0 - \cos(\omega t)/\omega$ (where A is the signal magnitude and $x_0$ is a constant determined by the initial conditions). The output equation can then be expressed as $y = g(A-\cos(\omega t)/\omega)(-\omega^2 A\sin(\omega t))$. It is evident that this has a similar form and properties as a memristive system with the exception of the frequency response from the $\omega^2$ term and that the hysteresis loop is in the first and third quadrants only for negative functions $g(x)$. In contrast to memristive systems this system does not degenerate to a linear resistor at high frequencies.

Example 8: nth order memresistor.

$$y = g(\boldsymbol{x}, u)\frac{d^{2m}u}{dt^{2m}}$$
$$\frac{d\boldsymbol{x}}{dt} = f(\boldsymbol{x}, u) \quad (10)$$

This example is a generalization of the previous example. As before $u$ represents the input signal, $\boldsymbol{x}$ is a state vector, and $y$ is the output signal. The function $g(\boldsymbol{x},u)$ is a continuous scalar function and $f(\boldsymbol{x},u)$ is a continuous vector function while $m$ is a positive integer (i.e. 1,2,3..). It is evident that this has a similar form and properties as a memristive system including zero-crossing hysteresis and exhibits zero-crossing hysteresis in the first and fourth quadrants for even m and positive functions $g(x,u)$. This system may or may not degenerate to a linear resistor at high frequencies depending on the form of the functions $f(\boldsymbol{x},u)$ and $g(\boldsymbol{x},u)$.

Example 9: nth order memresistor.

$$y = g_0(\boldsymbol{x},u)u + g_1(\boldsymbol{x},u)\frac{d^2u}{dt^2} + g_2(\boldsymbol{x},u)\frac{d^4u}{dt^4}\cdots$$
$$\frac{d\boldsymbol{x}}{dt} = f(\boldsymbol{x},u) \quad (11)$$

This example is a further generalization of the previous example expressed as a series expansion and including a memristive system as the first term. As before $u$ represents the input signal, $\boldsymbol{x}$ is a state vector, and $y$ is the output signal. The functions $g_0(\boldsymbol{x},u)$, $g_1(\boldsymbol{x},u)$, $g_2(\boldsymbol{x},u)$, etc. are continuous scalar functions and $f(\boldsymbol{x},u)$ is a continuous vector function. It is evident that this has a similar form and properties as a memristive system including zero-crossing hysteresis. This system may or may not degenerate to a linear resistor at high frequencies depending on the form of the *f(x,u)* and the *g(x,u)* functions.

Example 10: 3$^{rd}$ order memresistor.

$$y = g(x)(w(t) - w_0)$$
$$\frac{dx}{dt} = u \quad (12)$$
$$\frac{d^2(w(t)-w_0)}{dt^2} = \omega^2 u$$

In this example, *u* represents the input signal, *x* and *w* are state variables, $w_0$ is the initial condition of the state *w*, and *y* is the output signal with *g(x)* representing a continuous function of the system state *x*. For $u = A\sin(\omega t)$, $x(t) = A'\text{-}A\cos(\omega t)/\omega$ (where A and A' are constants) and $w(t)\text{-}w_0 = \omega^2(B+Ct) - A\sin(\omega t)$. Given initial conditions $w(0) = w_0$ and $dw(0)/dt = -\omega A$ this can be simplified to $w(t)\text{-}w_0 = -A\sin(\omega t)$. The output equation can then be expressed as $y = g(A'\text{-}A\cos(\omega t)/\omega)(-A\sin(\omega t))$ which exhibits zero-crossing hysteresis in the first and third quadrants for negative functions *g(x)*.

Example 11: (n+2)$^{th}$ order memresistor.

$$y = g(x, w, u, \frac{du}{dt})(w(t) - w_0)$$
$$\frac{dx}{dt} = f(x, w, u) \quad (13)$$
$$\frac{d^2(w(t)-w_0)}{dt^2} = \omega^2 u$$

This is a generalization of the previous example in which *u* represents the input signal, *x* is an n-dimensional state vector, *w* is a state variable, $w_0$ is the initial condition of the state *w*, and *y* is the output signal. The function *g(x,w,u,du/dt)* represents a continuous scalar function of the system state *x*, *w* and the input *u* and its derivative *du/dt*. The function *f(x,w,u)* represents a continuous vector function of the state *x*, *w* and the input *u*. For $u = A\sin(\omega t)$, $w(t)\text{-}w_0 = \omega^2(B+Ct) - A\sin(\omega t)$. Given initial conditions $w(0) = w_0$ and $dw(0)/dt = -\omega A$ this can be simplified to $w(t)\text{-}w_0 = -A\sin(\omega t)$. The output signal *y* is then in phase with the input signal *u* and zero-crossing hysteresis is produced in the *y* vs. *u* curve.

Example 12: 2$^{nd}$ order memresistor.

$$y = g(x)(w(t) - w_0)$$
$$\frac{dx}{dt} = u \quad (14)$$
$$\frac{d(w(t)-w_0)}{dt} = u^2 - \frac{1}{2}A^2$$

In this example, *u* represents the input signal, *x* and *w* are state variables, $w_0$ is the initial condition of the state *w*, and *y* is the output signal with *g(x)* representing a continuous function of the system state *x*. For $u = A\sin(\omega t)$, $x(t) = A'\text{-}A\cos(\omega t)/\omega$ (where A and A' are constants) and $w(t)\text{-}w_0 = -A^2\sin(2\omega t)/4\omega$. The output equation can then be expressed as $y = g(A\text{-}\cos(\omega t)/\omega)(-A^2\sin(2\omega t)/4\omega)$ which exhibits zero-crossing hysteresis in the first and third quadrants for negative functions *g(x)*.

Example 13: (n+1)$^{th}$ order memresistor.

$$y = g(x, w, u, \frac{du}{dt})(w(t) - w_0)$$
$$\frac{dx}{dt} = f(x, w, u) \quad (15)$$
$$\frac{d(w(t)-w_0)}{dt} = u^2 - \frac{1}{2}A^2$$

This is a generalization of the previous example in which *u* represents the input signal, *x* is an n-dimensional state vector, *w* is a state variable, $w_0$ is the initial condition of the state *w*, and *y* is the output signal. The function *g(x,w,u,du/dt)* represents a continuous scalar function of the system state *x*, *w* and the input *u* and its derivative *du/dt*. The function *f(x,w,u)* represents a continuous vector function of the state *x*, *w* and the input *u*. For $u = A\sin(\omega t)$, $x(t) = A'\text{-}A\cos(\omega t)/\omega$ (where A and A' are constants) and $w(t)\text{-}w_0 = -A^2\sin(2\omega t)/4\omega$. The output signal *y* is then in phase with the input signal *u* and zero-crossing hysteresis is produced in the *y* vs. *u* curve.

Example 14: n$^{th}$ order memresistor.

$$y = g(x, u) - g(x, 0)$$
$$\frac{dx}{dt} = f(x, u) \quad (16)$$

This is a generalization of the physical examples from examples 5 and 6. As in previous examples $u$ represents the input signal, $x$ is an n-dimensional state vector, and $y$ is the output signal. The function $g(x,u)$ represents a continuous scalar function of the system state $x$ and the input $u$ and $f(x,u)$ represents a continuous vector function of the state $x$ and the input $u$. The output signal $y$ is then in phase with the input signal $u$ since $u=0$ implies $y=0$ and zero-crossing hysteresis is produced in the $y$ vs. $u$ curve. It is notable that some memristive systems are special cases of this example for $g(x,u) = g_1(x)(u+x)$.

It is noted that it is unlikely that the above cases are exhaustive of all types of memresistive systems and professional mathematicians with expertise in non-linear differential equations would likely be able to point to other examples which may be relevant to the analysis of memory resistors. In addition a linear combination of the output signals $y$ in the previous 8 examples would also exhibit zero-crossing hysteresis allowing for some application of the principle of superposition. Thus even when a pinched hysteresis curve is in the second and fourth quadrant it may be convertible to a memresistor in the first and third quadrant by superposition with another pinched hysteresis curve in the first and fourth quadrant.

The last example illustrates that zero-crossing hysteresis can also be produced by memory-less dynamic systems.

Example 15: Non-linear differential equation.

$$y = \left(1 - \frac{du}{dt}\right) u \qquad (17)$$

For $u = A\sin(\omega t)$, $y = (1-\omega A\cos(\omega t)) A\sin(\omega t)$. In this case the $y$ vs. $u$ curve has the same form as examples 1 and 2 for $\omega=1$ and $A=1$. Since no integrals are involved this establishes that memory is not even necessary for a zero-crossing hysteresis curve to exist.

## III. CONCLUSION

The examples in this paper illustrate that dynamic systems outside of the canonical formulation of Chua and Kang's 1976 paper are capable of producing pinched hysteresis curves. Thus, even given the recently revised definition of a memristor, the claims of Chua and Williams that all resistance memory are memristors have no evident merit. It is suggested that *memresistor* be used as a more generic term to refer to any 2-terminal device exhibiting a memory resistance effect in order to distinguish from the original "4[th] fundamental circuit element" memristor definition relating electric charge and magnetic flux linkage.

For further reasons that the "memristor" is not a valid term for all memory resistors the reader is referred to [13].

Example 1: y = u+2x, dx/dt = u^2 - 0.5

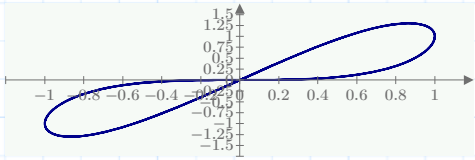

$sin(t) - 0.5 \cdot sin(2 \cdot t)$ (1)

$sin(t)$ (1)

Example 2: y = u + x du/dt, d2x/dt2 = u

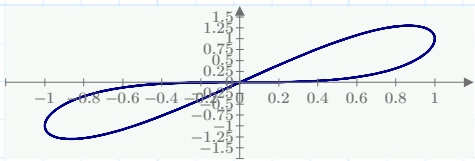

$sin(t) - sin(t) \cdot cos(t)$ (1)

$sin(t)$ (1)

Example 3: y = u + (1-x^2) x, dx/dt = u

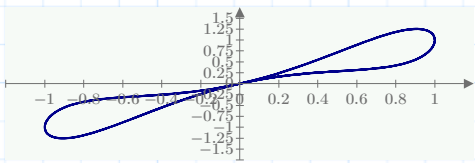

$sin(t) + (sin(t))^2 \cdot cos(t)$ (1)

$sin(t)$ (1)

## Case 1: Filamentary memresistor with nominal ion level nf

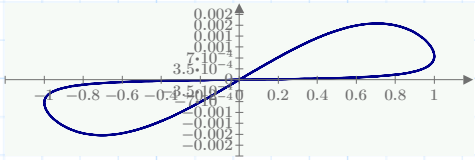

$$e^{-3\cdot\sqrt{(1+0.5\cdot\cos(x))^3 - 0.01\cdot\sin(x)\cdot(1+0.5\cdot\cos(x))^2}} - e^{-3\cdot\sqrt{(1+0.5\cdot\cos(x))^3}} \quad (1)$$

$$\underline{\sin(x)} \quad (1)$$

## Case 2: Filamentary memresistor with ion level 10*nf

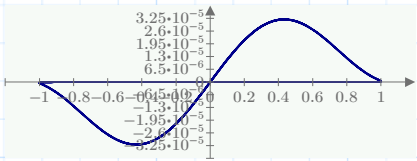

$$e^{-3\cdot\sqrt{10\cdot(1+0.5\cdot\cos(x))^3 - 0.01\cdot\sin(x)\cdot(1+0.5\cdot\cos(x))^2}} - e^{-3\cdot\sqrt{10\cdot(1+0.5\cdot\cos(x))^3}} \quad (1)$$

$$\underline{\sin(x)} \quad (1)$$

## Case 3: Filamentary memresistor with ion level nf/10

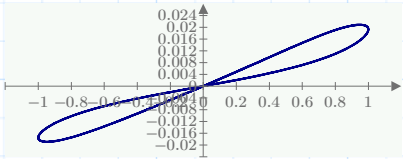

$$e^{-3\cdot\sqrt{0.1\cdot(1+0.5\cdot\cos(x))^3 - 0.01\cdot\sin(x)\cdot(1+0.5\cdot\cos(x))^2}} - e^{-3\cdot\sqrt{0.1\cdot(1+0.5\cdot\cos(x))^3}} \quad (1)$$

$$\underline{\sin(x)} \quad (1)$$

## Case 4: Filamentary memresistor with phase shift

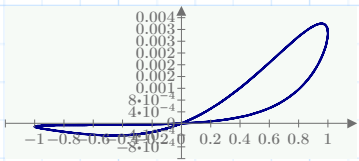

$$e^{-3\cdot\sqrt{(1+0.5\cdot\cos(x+1))^3 - 0.01\cdot\sin(x)\cdot(1+0.5\cdot\cos(x+1))^2}} - e^{-3\cdot\sqrt{(1+0.5\cdot\cos(x+1))^3}} \quad (1)$$

$$\underline{\sin(x)} \quad (1)$$